\documentclass[sigconf]{acmart}

\usepackage{multirow}
\usepackage{array}
\usepackage{enumitem}
\usepackage{CJKutf8}
\usepackage{graphicx} 
\usepackage{xcolor}

\copyrightyear{2024}
\acmYear{2024}
\setcopyright{acmlicensed}
\acmConference[ASE '24]{39th IEEE/ACM International Conference on Automated Software Engineering }{October 27-November 1, 2024}{Sacramento, CA, USA}
\acmBooktitle{39th IEEE/ACM International Conference on Automated Software Engineering (ASE '24), October 27-November 1, 2024, Sacramento, CA, USA} \acmDOI{10.1145/3691620.3695023}
\acmISBN{979-8-4007-1248-7/24/10}

\begin{document}

\title[AVIATE: Improving Traceability in Bilingual Projects]{AVIATE: Exploiting Translation Variants of Artifacts to Improve IR-based Traceability Recovery in Bilingual Software Projects}







\author{Kexin Sun}
\affiliation{
  \institution{State Key Lab for Novel Software Technology, Nanjing University}
  \city{Nanjing}
  \country{China}
}
\email{kexinsun@smail.nju.edu.cn}

\author{Yiding Ren}
\affiliation{
  \institution{State Key Lab for Novel Software Technology, Nanjing University}
  \city{Nanjing}
  \country{China}
}
\email{ryd_1@outlook.com}

\author{Hongyu Kuang}
\authornote{Corresponding author}
\affiliation{
  \institution{State Key Lab for Novel Software Technology, Nanjing University}
  \city{Nanjing}
  \country{China}
}
\email{khy@nju.edu.cn}

\author{Hui Gao}
\affiliation{
  \institution{State Key Lab for Novel Software Technology, Nanjing University}
  \city{Nanjing}
  \country{China}
}
\email{ghalexcs@gmail.com}

\author{Xiaoxing Ma}
\affiliation{
  \institution{State Key Lab for Novel Software Technology, Nanjing University}
  \city{Nanjing}
  \country{China}
}
\email{xxm@nju.edu.cn}

\author{Guoping Rong}
\affiliation{
  \institution{State Key Lab for Novel Software Technology, Nanjing University}
  \city{Nanjing}
  \country{China}
}
\email{ronggp@nju.edu.cn}

\author{Dong Shao}
\affiliation{
  \institution{State Key Lab for Novel Software Technology, Nanjing University}
  \city{Nanjing}
  \country{China}
}
\email{dongshao@nju.edu.cn}

\author{He Zhang}
\affiliation{
  \institution{State Key Lab for Novel Software Technology, Nanjing University}
  \city{Nanjing}
  \country{China}
}
\email{hezhang@nju.edu.cn}


\begin{abstract}
Traceability plays a vital role in facilitating various software development activities by establishing the traces between different types of artifacts (e.g., issues and commits in software repositories).
Among the explorations for automated traceability recovery, the IR (Information Retrieval)-based approaches leverage textual similarity to measure the likelihood of traces between artifacts and show advantages in many scenarios.
However, the globalization of software development has introduced new challenges, such as the possible multilingualism on the same concept (e.g., ``\begin{CJK*}{UTF8}{gbsn}属性\end{CJK*}'' vs. ``attribute'') in the artifact texts, thus significantly hampering the performance of IR-based approaches.
Existing research has shown that machine translation can help address the term inconsistency in bilingual projects.
However, the translation can also bring in synonymous terms that are not consistent with those in the bilingual projects (e.g., another translation of ``\begin{CJK*}{UTF8}{gbsn}属性\end{CJK*}'' as ``property'').
Therefore, we propose an enhancement strategy called AVIATE that exploits translation variants from different translators by utilizing the word pairs that appear simultaneously across the translation variants from different kinds artifacts (a.k.a. consensual biterms). We use these biterms to first enrich the artifact texts, and then to enhance the calculated IR values for improving IR-based traceability recovery for bilingual software projects.
The experiments on 17 bilingual projects (involving English and 4 other languages) demonstrate that AVIATE  significantly outperformed the IR-based approach with machine translation (the state-of-the-art in this field) with an average increase of 16.67 in Average Precision (31.43\%) and 8.38 (11.22\%) in Mean Average Precision, indicating its effectiveness in addressing the challenges of multilingual traceability recovery.

\end{abstract}







\maketitle

\section{Introduction}
\label{chap:intruduction}

\begin{figure*}[t]
	\centering
\includegraphics[width=0.80\textwidth]{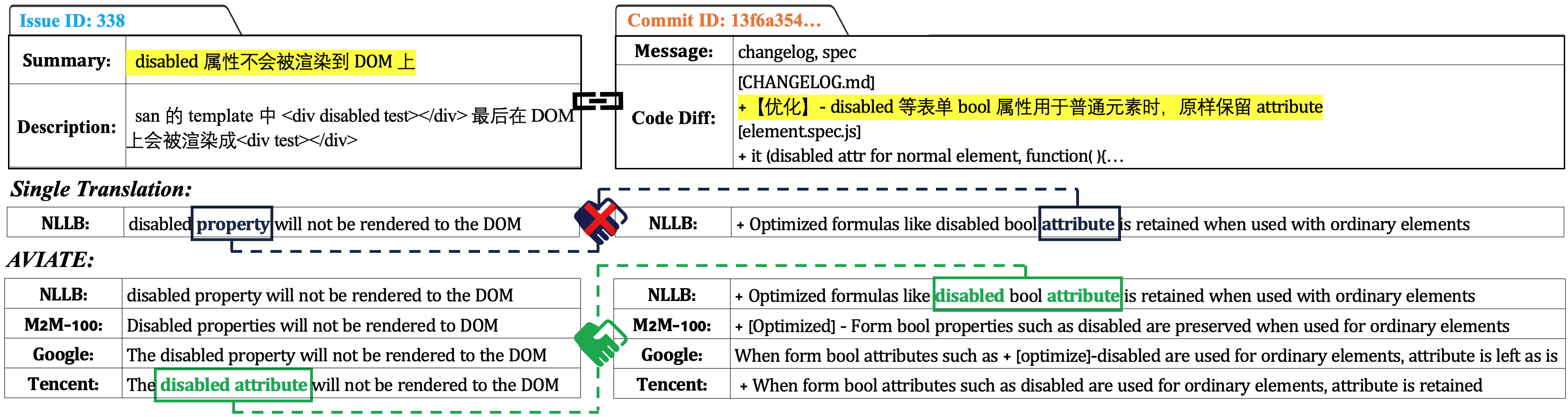} 
	\caption{Motivating example adapted from the San project.}
	\label{fig:motivation}
\end{figure*}

Software traceability is defined as ``the ability to interrelate any uniquely identifiable software engineering artifact to any other, maintain required links over time, and use the resulting network to answer questions of both the software product and its development process''~\cite{CoEST}.
It can support a wide range of software engineering (SE) activities, including facilitating the system comprehension~\cite{DBLP:conf/iceis/McharfiADBK15,DBLP:conf/seke/FilhoZ17}, accelerating the maintenance tasks~\cite{DBLP:journals/ese/MaderE15,DBLP:conf/euromicro/CharalampidouAC18}, managing the change process~\cite{rejab2019fuzzy,DBLP:journals/isse/KchaouBMB19}, and supporting daily development~\cite{DBLP:journals/tosem/ZhengCT18,DBLP:conf/iccip/RubasingheMP17}.
Unfortunately, despite its demonstrated benefits across many SE domains, recovering and maintaining traceability is time-consuming and challenging in practice~\cite{DBLP:conf/re/EgyedGG10}.
To handle this, researchers have tried various techniques, such as information retrieval (IR)~\cite{DBLP:conf/kbse/GaoKSMEMRSZ22,gao2024triad} and machine learning (ML)~\cite{DBLP:journals/jzusc/DuSHYW20,DBLP:conf/icse/LinLZ0C21}, to propose the approaches of automated traceability.
Among their explorations, the IR techniques, which can function effectively without any pre-existing traced data, become the mainstream technique for traceability recovery in current research and practice.
The IR models (such as Vector Space Model (VSM)~\cite{DBLP:journals/tse/AntoniolCCLM02}, Latent Semantic Indexing (LSI)~\cite{DBLP:conf/icse/MarcusM03}, and Jenson-Shannon model (JS)~\cite{DBLP:conf/iwpc/AbadiNS08}) primarily measure the correlation between different types of artifacts (e.g., issues and commits in online software repository) based on their textual similarity, where two given artifacts sharing more similar and meaningful terms (describing the system functionalities) are more likely to be relevantly traced.

Previous works in this field have largely assumed that all of the artifacts are written in a single language (i.e., English).
However, with the increasing globalization of software development, this assumption is no longer tenable.
Although English remains the primary programming and communication language in the global development activities, many non-native English-speaking developers tend to intermix their native language with English during their daily development tasks, such as committing issues or composing code comments~\cite{DBLP:conf/icgse/Lutz09,DBLP:conf/msr/LiuLC20}.
An analysis~\cite{DBLP:conf/lats/PiechA20} conducted on 1.1 million GitHub users revealed that non-English coding is a large-scale phenomenon.
They found that 12.7\% of the analyzed users wrote commit messages in non-English languages.
Of those, Chinese was the most common (28.6\% of non-English committers), followed by Spanish, Portuguese, French, and Japanese.
More than 100 kinds of languages were detected in commit messages on public Java projects.
The multilingualism in artifacts thus poses additional challenges for IR-based traceability recovery, especially for the notorious \textit{vocabulary mismatch problem}. 
This problem says that the performance of IR-based approaches can be greatly hindered when different terms are used to denote the same concept, and our discussed multilingualism in SE makes the situation even worse.

Recently, the work by Lin et al. ~\cite{DBLP:journals/ese/LinLC22} has pioneered the exploration of using machine translation as a potential solution for addressing multilingualism in traceability recovery.
They compared the performance of translative IR-based approaches (i.e. first translating the artifacts into the same language—English, and then using IR models to recover traceability on the translated artifacts) and multilingual BERT-based learning approaches for recovering the trace links between issues and commits on 17 bilingual projects (14 English-Chinese projects and 3 English-other-language projects). 
They found that the translative IR-based approaches generally perform better on the traceability recovery among intra-project bilingual artifacts, which represents a current state-of-the-art (SOTA) approach in multilingual traceability recovery.
However, despite the promising results of using machine translation, the translation inconsistency issue still remains and undermines the performance of IR-based approaches.
For example, the term ``\begin{CJK*}{UTF8}{gbsn}属性\end{CJK*}'' from an issue could be translated as either ``attribute'' or ``property'' by the machine translator.
These two translations are both reasonable, but which one is better for the IR-based approach depends on the corresponding word (may also be translated) used in the relevant commit message, i.e., both the issue and the commit message use ``attribute'', or vice versa.
Such inconsistencies are basically irrelevant to the performance of machine translator because they are not aware of the multilingual project context during training.
However, the IR-based approaches should deal with the translation-inconsistency-issue because they will process the texts from different kinds of artifacts, and their performance will be further improved for automated traceability recovery on multilingual software projects (further discussions on the translation-inconsistency-issue by a motivation example is in Section~\ref{chap:motivating_example}).

\begin{figure*}[t]
	\centering
\includegraphics[width=0.94\textwidth]{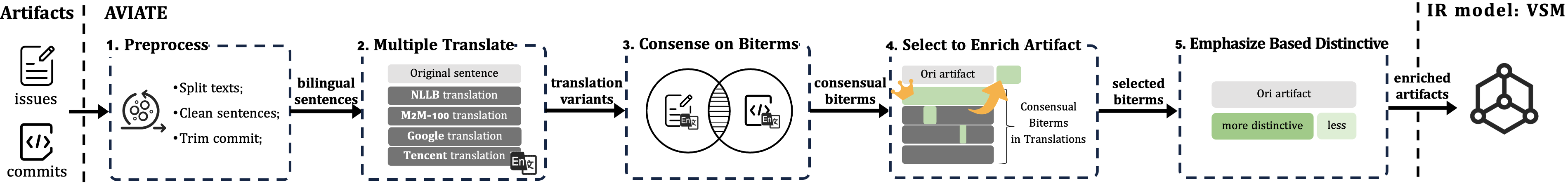} 
	\caption{Overview of the AVIATE framework.}
	\label{fig:overview}
\end{figure*}

Hence, to mitigate the potential inconsistency in translating terms and fully leverage the benefits of machine translation, we decide to ``turn lemons into lemonade’’, i.e., to exploit translation variants from multiple SOTA translators to thoroughly capture the implicit system functionalities described in multilingual artifacts.
In particular, we use the ``consensual biterm’’ proposed and utilized by Gao et al. ~\cite{DBLP:conf/kbse/GaoKSMEMRSZ22,gao2024triad} for enhancing IR-based traceability.
These biterms are the word pairs that simultaneously appear in two different types of artifacts to indicate at least part of the same system functionalities that are characterized in different software artifacts.
Following and adapting this idea, we first generate candidate biterms from multiple translation variants (i.e., the different translated versions of the same sentence) within each kind of artifact (i.e., issues and commits in this paper), and then use these candidate biterms (augmented by translation variants for each artifact) to extract consensual biterms as the basis of our following enhancing strategies.
We propose our approach named \textbf{AVIATE} (tr\textbf{A}nslation \textbf{V}ariant b\textbf{I}lingu\textbf{A}l \textbf{T}raceability r\textbf{E}covery) to improve the IR-based traceability recovery for the multilingual artifacts (mainly focusing on the most common ones where developers mix English with his native language in bilingual software projects).
Specifically, we first employ four mainstream translators (i.e. NLLB-1.3B model~\cite{DBLP:journals/corr/abs-2207-04672}, M2M-100-12B model ~\cite{DBLP:journals/jmlr/FanBSMEGBCWCGBL21}, Google Translate~\cite{Google}, and Tencent Translate~\cite{Tencent}) to generate translation variants for sentences (with necessary pre-processing) in the artifacts that are not purely in English.
We then employ the natural-language-processing tool CoreNLP~\cite{CoreNLP} to traverse translation variants and generate candidate biterms.
By intersecting the biterms derived from the two kinds of artifacts, we obtain the initial consensual biterms.
Because these consensual biterms are extracted from a much larger corpus that is augmented by translation variants (i.e., each bilingual sentence in a give artifact will be expanded to four translate variants in this paper), we further proposed a TF-IDF-like metric for our filtering strategy to select \textit{distinctive consensual biterms} as the final enrichment to improve the quality of bilingual artifact texts.
We argue that \textit{these distinctive consensual biterms extracted from translation variants consistently capture the vital descriptions of system functionalities across the bilingual artifacts, thus being important to improve IR-based traceability in bilingual software projects}.
Accordingly, we then proposed an enhanced weighting factor for the selected biterms to ensure that the important information captured by distinctive consensual biterms receives adequate attention in the IR model.
Finally, we use the VSM, a simple, effective, and robust IR model which has been validated in many traceability researches (e.g., ~\cite{DBLP:conf/sigsoft/LoharAZC13,DBLP:journals/ese/LinLC22}), to calculate the textual similarity between the enriched and weighting-factor-enhanced artifacts and generate the candidate trace list accordingly.

To validate the effectiveness of AVIATE, we conducted experiments on the dataset containing 17 bilingual projects proposed by Lin et al.~\cite{DBLP:journals/ese/LinLC22}.
First, we evaluated the performance of VSM based on the translations from four different translators and identified an optimal single-translation-based IR approach (i.e., VSM based on Tencent Translate service). 
Then, we utilized AVIATE to enhance the identified best translation results with the consensual biterms extracted from multiple translation variants. 
Additionally, we introduced the consensal-biterm-based enhancement strategy TAROT proposed by Gao et al.~\cite{DBLP:conf/kbse/GaoKSMEMRSZ22} for comparison on the same bilingual projects, except that TAROT does not take multiple translations into account.
Experimental results indicate that compared to TAROT, AVIATE can provide a more significant improvement to the best Tencent-Translate-based VSM.
AVIATE increases the average AP (Average Precision) by 16.67 (a 31.43\% improvement), and achieves a maximum AP increase of 30.36 (101.17\%, the Canal project).
For MAP (Mean Average Precision), AVIATE achieves an average improvement of 8.38 (11.22\%), and the maximum MAP improvement of 19.45 (26.64\%, the Rax project).
AVIATE also outperforms TAROT by 8.46 (13.81\%) in AP and 3.17 (3.96\%) in MAP.
Therefore, we argue that AVIATE can serve as a robust solution to address the challenges posed by multilingualism in artifact traceability.
In summary, the main contributions of this work are: (i) a novel approach called AVIATE exploiting translation variants of artifacts to improve IR-based traceability recovery in bilingual software projects; (ii) an empirical evaluation of TRIAD on seventeen open-source systems; and (iii) availability of source code and data of AVIATE at https://github.com/huiAlex/AVIATE.

\section{Motivating Example}
\label{chap:motivating_example}
In this section, we present an example adapted from the English-Chinese project “San” (a JavaScript component framework proposed by the Chinese IT Company Baidu, for details please refer to Section \ref{chap:exp_dataset}) to demonstrate the challenges posed by the term translation inconsistency in the traceability recovery on bilingual projects. 
Figure~\ref{fig:motivation} shows a traced pair of issue and commit in this project.

In the figure, the issue (Issue 338) consists of a summary and a description, and the commit (Commit 13f6a354...) consists of a message and a code diff (where lines starting with ``+'' indicate additions or modifications, and lines starting with ``-'' indicate deletions). 
The highlighted parts in the figure emphasize the statements in two artifacts that reflect their traceability relationships.
Specifically, to address the issue of ``\begin{CJK*}{UTF8}{gbsn}disabled \textbf{属性}不会被渲染到 DOM 上\end{CJK*}'' (\textit{The ``disable’’ \textbf{attribute} not being rendered to the DOM}), developers optimized the handling to the disabled \textbf{attribute} by modifying \texttt{CHANGELOG.md} and \texttt{element.spec.js} files.
When we use NLLB~\cite{DBLP:journals/corr/abs-2207-04672} (a SOTA translator, further discussed in Section~\ref{chap:approach_step2}) to translate both of the artifacts, the term ``\begin{CJK*}{UTF8}{gbsn}属性\end{CJK*}'' is translated inconsistently, appearing as ``\underline{property}'' in the issue and as ``\textbf{attribute}'' in the commit. 
This inconsistency results in a semantic mismatch, thus hindering the IR model from capturing the traceability link between the two artifacts.
Even though both of them have been translated into the same language. 

However, by applying our proposed approach AVIATE, we can successfully identify the distinctive consensual biterm \textit{(disabled, attribute)} from each four translation variants of the issue and the commit.
This extracted biterm can both captures the essential concept shared between two artifacts, and utilizes it to enrich the relatively short translated texts of artifact.
Thus, AVIATE can bridge the semantic gap introduced by inconsistent translations, and enable the IR model to recover the traceability link with better performance.

\section{Proposed Approach}

The overview of our proposed AVIATE framework is shown in Figure \ref{fig:overview}.
Specifically, we first pre-process the artifacts (Section~\ref{chap:approach_step1}). 
Then, we utilize four mainstream translators to translate non-English sentences within the artifacts (Section~\ref{chap:approach_step2}).
Third, we extract consensual biterms from the multiple translations of issues and commits (Section~\ref{chap:approach_step3}).
Finally, we further select distinctive consensual biterms to enrich the artifact text (Section~\ref{chap:approach_step4}), and adjust their weighting factor to emphasize the critical sections (Section~\ref{chap:approach_step5}).
We employ the IR model, VSM, to compute the similarity between enriched artifacts and generate a candidate trace list (Section~\ref{chap:approach_step6}).

\subsection{Pre-processing Data}
\label{chap:approach_step1}
We pre-process the text content of input artifacts, including issue summary, issue description, commit message, and commit diff, as follows:
\textbf{1) First}, we split the text into sentences based on characters, such as ``\textbackslash n'', ``\textbackslash r''.
\textbf{2) Then}, we clean the sentences by removing URLs, file paths, personal mentions (e.g., ``@xxx''), formatting characters (e.g., ``**'' for bold), and irrelevant special punctuations.
\textbf{3) Lastly}, we trim the content of commit diffs.
Because we found that the length of commit diffs significantly exceeds that of the commit message (see related ratio in Table~\ref{tab:statistics}).
This results in the important summaries within the commit messages being overshadowed by the detailed code changes in the commit diffs, and consequently, they cannot receive sufficient attention in the IR model.
To handle this, we trim less important changes in the commit diffs.
We remove all changes to configuration files (e.g., ``.properties''/``.pom'' files), template files (e.g., ``.axml''/``.vue'' files), and style files (e.g., ``.css''/``.styl'' files), as these are mostly generated by development tools.
For changes to other types of files, we only retain the added content (in the line starting with ``+''), which represents the primary work conducted by the developer.

\subsection{Translating with Multiple Translators}
\label{chap:approach_step2}
To mitigate the potential term translation inconsistency which has been exemplified in Figure~\ref{fig:motivation}, we introduce four mainstream translators to complete the translation task.
They are two Many-to-Many multilingual translation models (\textbf{NLLB-1.3B model}~\cite{DBLP:journals/corr/abs-2207-04672} and \textbf{M2M-100-12B model}~\cite{DBLP:journals/jmlr/FanBSMEGBCWCGBL21} and two web-based translation services (\textbf{Google Translate}~\cite{Google} and \textbf{Tencent Translate}~\cite{Tencent}). 
Among them, the NLLB-1.3B model is designed for low-resource natural language training and can support translation between 202 different languages.
The M2M-100-12B model is a non-English-Centric translation model and can translate between 100 languages.
Both of them have been widely used in various research domains~\cite{DBLP:journals/corr/abs-2304-04675}.
On the other hand, Google Translate and Tencent Translate are two widely used web-based translation services known for their accuracy and efficiency.
Wu et al.'s research~\cite{DBLP:journals/corr/WuSCLNMKCGMKSJL16} indicates Google Translation has addressed approximately 60\% of previously known translation errors in popular languages, significantly enhancing translation performance.
Tencent Translate has also demonstrated outstanding performance in English-to-Chinese translation~\cite{DBLP:conf/wmt/WuWWLXTSL20,DBLP:conf/wmt/WangLL0T0WZZ21}.

We use regular expressions to identify sentences that contain non-English characters, and translate them into pure English sentences using the above four translators.
We refer to the different translation versions for the same sentence as \textit{Translation Variants}.
In our approach, each original non-pure-English sentence corresponds to four translation variants.
The pure English texts can be considered as a special case where there is only one variant.
Through this translation process, we can obtain a pure English corpus.
We denote the corpus obtained from issues as $TV_{issues}$ and the corpus obtained from commits as $TV_{commits}$.

\begin{figure}[t]
	\centering
	\includegraphics[width=0.38\textwidth]{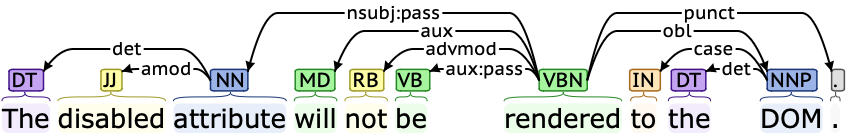} 
	\caption{Example of dependency graph parsed by CoreNLP.}
	\label{fig:dep_example}
\end{figure}

\subsection{Extracting Consensual Biterms from Artifact Translations}
\label{chap:approach_step3}
Next, we extract consensual biterms from the translation variants of issues and commits.
These consensual biterms not only represent the alignment of system functionalities described in two kinds of artifacts, but also ensure the term consistency in translation.
We utilize Stanford CoreNLP~\cite{CoreNLP} to help us parse the syntactic structures of artifact texts and extract biterms.
The specific process is as follows:
\textbf{1) First}, we tag each term’s Parts-Of-Speech (POS) in the variant sentence, such as nouns, verbs, adjectives, and determiners.
\textbf{2) Second}, we extract candidate biterms based on the dependency parsing results from CoreNLP.
As shown in Figure~\ref{fig:dep_example}, CoreNLP's dependency parsing can help us extract pairs of terms in the sentence that have syntactic relationships.
We use ``{\small$Gov\xrightarrow{Reln}Dep$}'' to describe a pair of syntactic-related terms in the dependency graph, where $Gov$ denotes the dominant term, $Dep$ denotes the dependent term, and $Reln$ denotes the dependency relationship between them.
For example, ``{\small attribute$\xrightarrow{amod}$disabled}'' represents the term ``disabled'' ($Dep$) is an adjectival modifier ($Reln=$``$amod$'') of the term ``attribute'' ($Gov$).
We traverse each syntactic-related pair in the dependency graph, and only retain those where both $Gov$ and $Dep$ are nouns, verbs, or adjectives as candidate biterms.
Because previous studies~\cite{DBLP:journals/infsof/AliCHH19,DBLP:conf/iwpc/LuciaPOPP11} have suggested that compared to other POSs, these three types of POSs are more likely to carry important semantics for the artifacts.
Therefore, we remove the biterms, such as (``attribute’’, ``the’’, ``det’’), from the final candidate biterms because the term ``the’’ is a determiner (POS=``DT’’) here, which makes such biterms express less semantics than those entirely composed of nouns, verbs, or adjectives, such as (``attribute’’, ``disabled’’, ``amod’’).
Here, We use the form ($Gov$,$Dep$,$Reln$) to denote a candidate biterm.
For $Gov$ and $Dep$, we use their lowercased stems form to represent the terms.
Additionally, we retain the syntactic relationship ($Reln$) in the candidate biterm, which allows us to further restrict the scope of consensus when matching them.
We denote the candidate biterms extracted from the variants in $TV_{issues}$ as $BT_{issues}$ and those extracted from the $TV_{commits}$ as $BT_{commits}$.
\textbf{3) Lastly}, we select the biterms that simultaneously appear in the candidate biterms extracted from issues and commits as consensual biterms. We denote the set of these consensual biterms as $ConsBT_{total}$, where $ConsBT_{total}=BT_{issues}\cap BT_{commits}$.

\subsection{Selecting Distinctive Consensual Biterms to Enrich Texts}
\label{chap:approach_step4}
For the consensual biterms from different translation variants, we further choose those more distinctive to enrich the artifact texts.
Referring to the concept of Inverse Document Frequency (IDF)~\cite{DBLP:books/aw/Baeza-YatesR99}, we propose Inverse Translation Variant Frequency (ITVF) to measure the distinctiveness of the consensual achieved by a biterm, which is defined as:
\begin{equation}
ITVF(bt) = 
\begin{cases}
  \log \left( \frac{N}{\hspace{0.4em}1\hspace{0.1em}+\hspace{0.1em} TVF(bt)\hspace{0.4em}} \right) , & \text{if $bt$ $\in$ $ConsBT_{total}$} \\
  \text{    0    } , & \text{if $bt$ $\notin$ $ConsBT_{total}$ }
\end{cases}
\end{equation}
where $bt$ represents a biterm, $N$ denotes the total number of translation variants in $TV_{issues}$ and $TV_{commits}$, $TVF(bt)$ stands for the Translation Variant Frequency (TVF), indicating the number of translation variants which contain the biterm $bt$.
In this formula, we consider a consensus that appears infrequently in the entire translation corpus as distinctive.
Since we are measuring the distinctiveness of the achieved consensus, we directly set the distinctiveness (i.e. $ITVF$ value) of non-consensual biterms to 0.

We calculate $ITVF$ value for each consensual biterm in $ConsBT_{total}$ and apply min-max normalization to scale their values to the range [0,1], which accounts for the differences in scale across various systems.
To ensure the comprehensiveness and distinctiveness of the consensual biterms selected, we adopt a two-tier selection process for each sentence.
For the sentence $s$, we first calculate the consensual distinctiveness score (denoted as $ConsDistinc$) for each of its translation variants, and select all consensual biterms from the highest-scoring translation variant for the subsequent text enrichment.
Given a translation variant $tv$, its consensual distinctiveness score can be calculated as:
\begin{equation}
ConsDistinc(tv) = \sum\nolimits_{i=1}^{\mid BT_{tv}\mid}  BTF(bt_{i})\cdot ITVF(bt_{i})
\end{equation}
where $BT_{tv}$ denotes the set of biterms extracted from variant $tv$, $bt_{i}$ is a biterm within $BT_{tv}$, $BTF(bt_{i})$ is the Biterm Frequency (BTF) of $bt_{i}$ in the translation variant $tv$, and $ITVF(bt_{i})$ represents the normalized distinctiveness score of $bt_{i}$.
This variant-level selection ensures the coverage of the selected consensual biterms.
To allow for more flexible selection and to magnify the advantages of multiple translations, we further select consensual biterms from the remaining translation variants at the biterm level.
We also consider consensual biterms with the $ITVF$ value greater than 0.6 as sufficiently distinctive, and select these biterms to complement the results obtained from the highest-scoring variant. 
This two-tier selection results in a refined set of consensual biterms for the sentence, denoted as $ConsBT_{s}$.
Gathering the consensual biterms from each sentence, we obtain the final consensual biterm set for the artifact $arft$, denoted as $ConsBT_{arft}$.

These consensual biterms represent a distilled compilation of multiple translations for the bilingual artifact contents.
By incorporating them into the artifact texts, we can enhance the correlation between artifacts within the IR model, especially for those that have achieved consensus.
For each consensual biterms in ConsBT$_{arft}$ (composed of ($Gov$,$Dep$,$Reln$)), we disregard the syntactic relationship ($Reln$) and concatenate the terms $Gov$ and $Dep$ in dictionary order into a string, then append it to the artifact.
For example, for (``render’’, ``dom’’, ``obl’’), we concatenate the terms ``render’’ and ``dom’’ in dictionary order as ``domrender'' and treat this concatenated result as a special term added to the end of the artifact text.

\begin{table*}[t]
  \footnotesize
  \centering
  \caption{Detailed information of bilingual projects in the evaluation datasets, including their language, affiliated company, domain description, and the numbers of issues, commits, and trace links, respectively.}
    \begin{tabular}{ccclccc}
    \toprule
    \textbf{Language} & \textbf{Project} & \textbf{Company} & \multicolumn{1}{c}{\textbf{Domain}} & \textbf{Issues } & \multicolumn{1}{l}{\textbf{Commits }} & \textbf{Links} \\
    \midrule
    \multirow{14}[1]{*}{\textbf{EN-CH}}
          & Arthas  & Alibaba  & Java diagnostics tool for production issues & 122   & 167   & 167 \\
          & bk-cmdb  & Tencent  & Enterprise-level configuration management system & 895   & 1,178  & 1,179 \\
          & Canal  & Alibaba  & MySQL Database log parser & 232   & 273   & 273 \\
          & Druid  & Alibaba  & Database connection pools in JAVA & 1,092  & 1,161  & 1,161 \\
          & Emmagee  & Netease  & Performance test tool for Android App & 31    & 32    & 32 \\
          & Nacos  & Alibaba  & Service discovery and management platform  & 132   & 161   & 161 \\
          & NCNN  & Tencent  & Neural network library for mobile computing  & 97    & 99    & 99 \\
          & Pegasus  & Xiaomi  & Distributed key-value storage system & 160   & 160   & 160 \\
          & QMUI Android  & Tencent  & Mobile UI library for Android & 70    & 71    & 71 \\
          & QMUI IOS  & Tencent  & Mobile UI for IOS & 32    & 35    & 35 \\
          & Rax   & Alibaba  & React framework for building application  & 560   & 571   & 571 \\
          & San   & Baidu  & JavaScript component framework  & 186   & 275   & 275 \\
          & Weui  & Tencent  & WeChat-like UI framework & 154   & 159   & 159 \\
          & xLua  & Tencent  & Lua library for integrating C\# & 52    & 52    & 52 \\
    \textbf{EN-KO}  & Konlpy & Personal  & NLP package for Korean & 32    & 33    & 33 \\
    \textbf{EN-JA}  & Cica  & Personal  & Font repository for Japanese  & 25    & 27    & 27 \\
    \textbf{EN-DE} & Aws-berline  & Personal  & Web application touring Berlin & 74    & 74    & 74 \\
    \bottomrule
    \end{tabular}%
  \label{tab:dataset}%
\end{table*}%

\subsection{Adjusting Weighting-Factor Based on Distinctiveness of Selected Biterms}
\label{chap:approach_step5}
Additionally, we adjust the adding count of the selected consensual biterms based on their distinctiveness scores to emphasize the importance of critical sections within the artifact.
Generally, we add each selected biterm to the artifact text once.
However, we observe that the information in the issue summaries and the commit messages is often concise but crucial.
If we treat this manually summarized information in the same way as the detailed descriptions in issue descriptions and commit diffs, these brief yet important information may not receive sufficient attention in the IR model.
Therefore, besides trimming the commit diffs during pre-processing, we further emphasize this information by increasing the occurences of the consensual biterms extracted from them according to their distinctive degrees.
(The semantics captured by the selected consensual biterms in the critical sections are the ``important information'' we emphasized in Section~\ref{chap:intruduction}.)
We denote the consensual biterms extracted from an issue summary or commit message as $ConsBT_{iss\_summ}$ and $ConsBT_{cmt\_msg}$ respectively, and determine their repetition counts according to the distinctiveness score of the issue summary or commit message, which can be expressed as:
\begin{equation}
ConsDistinc(\text{\footnotesize$iss\_summ$ or $cmt\_msg$}) = \frac{1}{|S|} \sum_{s \in S} ConsDistinc(s)
\end{equation}
\begin{equation}
ConsDistinc(s) = \max\nolimits_{tv \in TV_{s}} ConsDistinc(tv)
\end{equation}
where $ConsDistinc($iss\_summ$ or $cmt\_msg$)$ refers to the score of an issue summary or a commit message that is calculated as the average score of its constituent sentences (denoted as $S$). 
For each sentence $s$ in the set $S$, we select the highest score from all its translation variants to represent its consensual distinctiveness (i.e., $ConsDistinc(s)$).
To highlight the significance of this information relative to whole corpus, we further propose the Consensual Distinctiveness Emphasis Score, denoted as $EmphConsDistinc$.
We then utilize this score to determine the repetition count of their consensual biterms.
These two factors can be calculated as:
\begin{equation}
\begin{aligned}
    &EmphConsDistinc(\text{$iss\_summ$ or $cmt\_msg$}) = \\[-0.5em]
    &\max\left(ConsDistinc(\text{$iss\_summ$ or $cmt\_msg$}), \overline{ConsDistinc}\right)
\end{aligned}
\end{equation}
\begin{equation}
\text{$Repetition$ $Count$} = 0.1 \times EmphConsDistinc \times len(arft)
\end{equation}
where $\overline{ConsDistinc}$ represents the average consensual distinctiveness scores of all sentences in the corpus.
We use $\overline{ConsDistinc}$ as the lower bound for each issue summary and/or commit message to ensure these important information can consistently receive a certain degree of emphasis.
In the formula for $Repetition$ $Count$, ``0.1'' acts as a scaling factor for $EmphConsDistinc$.
We determined this parameter through a grid research in the range (0, 1) with a step size of 0.05, and found that 0.1 provides us a best overall performance (the benefits brought by the weighting-factor adjustment strategy significantly drop when the factor exceeds 0.3). 
This formula is to proportionately emphasize the important consensual biterms based on their distinctiveness and the length of the artifact text.

We now use the motivating example depicted in Figure~\ref{fig:motivation} to demonstrate how we emphasize important information in crucial sections by adjusting the weighting factors.
In this case, we focus on the consensual biterm ``attributedisabled'' appeared in the summary of Issue 338.
This biterm originates from the highest-scored translation variant (Tencent variant) of the sentence in the summary.
For Formula (1), the total number of translation variants is $N=25,316$ for the San project, and $TVF(attributedisabled)$ equals 2 for this biterm (the most unique one because only shared by two translation variants in the example), resulting in $ITVF(attributedisabled)$ as 3.93 (the highest consensus score and thus normalized as 1).
For Formula (2), we extracted three biterms from the Tencent variant: ``attributedisabled'', ``attributerender'', and ``domrender''.
Among these, only ``attributedisabled'' is a consensual biterm.
Therefore, we get $ConsDistinc(Tencent\,tv)=0.333$ ($BTF(attributedisabled)=0.333$, and $ITVF(attributedisabled)=1$).
Other variants contain no consensual biterms, resulting in $ConsDistinc(other\,tv)=0$.
As the summary of Issue 338 only consists of this sentence, we obtain $ConsDistinc(summary\,of\,Issue\,338)=ConsDistinc(s)=ConsDistinc(Tencent\,tv)=0.333$.
Similar to the above process, by considering 21,704 sentences in the San project, we eventually get $\overline{ConsDistinc}$ as 0.179.
Because $ConsDistinc(summary\,of\,Issue\,338)$ exceeds $\overline{ConsDistinc}$, we finally get $Repetition\,Count=1$ using $EmphConsDistinc=0.333$ and $len(Issue\,338)=15$ (after pre-processing).
Consequently, we insert ``attributedisabled'' to the end of the issue 2 times in total: once for routine and once for repetition.
(For ``attributedisabled'' in Commit 136f6a354...'s diff, we follow general processing and insert it once.)
This repetition process gives the newly added consensual biterm ``attributedisabled'' with the $TF$ value of $2/(2 + 15)\approx0.118$ in subsequent IR model calculations for Issue 338, which effectively mitigates semantic misalignment due to translation inconsistency by adding a new TF-IDF weighted biterm to Issue 338's document, thus enhancing the textual similarity between this pair of artifacts by approximately 10\% (0.3540$\to$0.3890).

\subsection{Generating Candidate Trace Links with VSM}
\label{chap:approach_step6}
We employ the IR model, VSM, to recover the traceability among the enriched artifacts, because it has achieved superior performance in many tracing scenarios~\cite{DBLP:conf/sigsoft/LoharAZC13,DBLP:journals/ese/LinLC22}.
The specific steps are outlined below:
\textbf{1) Creating and normalizing corpus}: We tokenize the enriched artifacts, converting them into lists of tokens (including both terms and biterms). For tokens adhering to the CamelCase naming convention, we further split them into multiple tokens. Then, we remove stop words from the token lists, convert the remaining tokens to lowercase, and apply the Porter stemming algorithm~\cite{DBLP:journals/program/Porter80} to preserve them in the stem form.
\textbf{2) Constructing artifact vectors and computing textual similarity}: We model each artifact as a TF-IDF vector, and calculate the textual similarity between two artifacts as the cosine of the angle between their vector representations.
\textbf{3) Ranking candidate links}: Each candidate trace link consists of an issue, a commit, and their textual similarity. 
We generate the candidate link list by sorting the issue-commit pairs in descending order of their textual similarity.

\section{Experiment Setup}
We conducted systematic experiments to assess the effectiveness of our IR-based enhancement strategy, AVIATE.
This section details our experimental setups. 

\subsection{Datasets}
\label{chap:exp_dataset}
We utilized the dataset proposed by Lin et al.~\cite{DBLP:journals/ese/LinLC22} to evaluate our approach and baseline approaches.
This dataset comprises issues, commits, and their trace links extracted from 17 popular open-source projects on GitHub.
These projects contain 14 English-Chinese (EN-CH) projects, one English-Korean (EN-KO) project, one English-Japanese (EN-JA) project, and one English-German (EN-DE) project.
To ensure bilingualism in the dataset, they removed the trace links composed of the same sole language, i.e., each trace link in the dataset now includes at least one associated artifact containing foreign language terms.
Lin et al. established the ground truth trace links between commits and issues by extracting explicitly defined issue tags from commits.
Since commits without explicitly mentioned issue tags are also prevalent, relying solely on explicit tags to establish trace links might lead to ``true positive'' instances being mistakenly treated as ``false positives'' in the evaluation step.
To partially mitigate this issue, they limited the scope of artifacts to those associated with at least one of the explicitly specified links in the ground truth link set.
This ensured a denser ground-truth link coverage and fewer inaccuracies.
Table~\ref{tab:dataset} describes the detailed information on these projects.
The released dataset did not preserve the structural information of the artifacts, including certain text formatting elements such as ``\textbackslash n'', and the distinction between different sections of the artifacts (i.e., issue summary or issue message, commit message or commit diff).
To better apply our approach, we re-crawled the corresponding artifact texts based on the issue and commit IDs provided in the dataset.
The reorganized dataset is also included in our replication package.

\subsection{Evaluation Metrics}
\label{chap:exp_metric}
We first introduce two metrics, precision and recall.
Precision represents the proportion of correct trace links among the retrieved candidate links, and recall represents the proportion of retrieved correct links among all true links.
They are defined as follows:
\begin{equation}
Precision = \frac{\hspace{1em}\mid\text{$retrieved$ $links$}\cap\text{$true$ $links$}\mid\hspace{1em}}{\mid\text{$retrieved$ $links$}\mid}
\end{equation}
\begin{equation}
Recall = \frac{\hspace{1em}\mid\text{$retrieved$ $links$}\cap\text{$true$ $links$}\mid\hspace{1em}}{\mid\text{$true$ $links$}\mid}
\end{equation}
where $retrieved$ $links$ represents the candidate link list retrieved by traceability recovery approaches, and $true$ $links$ represents the true links established through human.

Furthermore, following previous works~\cite{DBLP:journals/ese/LinLC22,gao2024triad}, we introduce two comprehensive metrics commonly used to access the performance of IR systems: average precision ($AP$) and mean average precision ($MAP$).
$AP$ is the average precision at each relevant document being retrieved.
It measures the quality of the ranking by considering both the precision and the order of retrieval.
$MAP$ is the average of the $AP$ scores of all queries.
It reflects the average precision level of the IR system when handling multiple queries.
$AP$ and $MAP$ are computed as:
\begin{equation}
AP = \frac{\hspace{1em}\sum_{r=1}^{N}{(Precision(r)\times isRelevant(r))}\hspace{1em}}{\mid\text{$true$ $links$}\mid}
\end{equation}
\begin{equation}
MAP = \frac{\hspace{1em}\sum_{q=1}^{Q}{AP(q)}\hspace{1em}}{Q}
\end{equation}
where $r$ is a rank in an ordered list of candidate links, $N$ is the total number of candidate links, $Precision(r)$ is the precision score at the cut-off rank $r$, and 
$isRelevant(r)$ is a binary function returning 1 if the link at rank 
$r$ is correct, otherwise it returns 0. 
For $MAP$, the $q$ is a single query, and $Q$ is the number of all queries.
In the scenario of traceability recovery, a query represents an issue within the project, and the retrieved document represents a commit.

\subsection{Research Questions}
\label{chap:exp_rq}
We use the following research questions to evaluate the improvement of AVIATE for the IR-based traceability recovery on bilingual projects, and the contributions of its different strategy settings:

\vspace{1mm}
\noindent
\textbf{RQ1: How much improvement can AVIATE bring to the single-translation-based IR approach?}

\vspace{1mm}
\noindent
\textbf{RQ2: How much contribution do the different strategy settings within AVIATE make?}


\vspace{1mm}

\begin{table*}[t]
  \footnotesize
  \centering
  \caption{The average score of computed AP, MAP evaluating the approaches (the basic VSM and 4 machine translation-based VSM) on different kinds of bilingual projects.
  ``()'' represents the number of projects of this type. 
  (RQ1)}
    \begin{tabular}{c|cc|cc|cc|cc|cc}
    \hline
    \multirow{2}[4]{*}{} & \multicolumn{2}{c|}{\textbf{$VSM$}} & \multicolumn{2}{c|}{\textbf{$VSM_{NLLB}$}} & \multicolumn{2}{c|}{\textbf{$VSM_{M2M-100}$}} & \multicolumn{2}{c|}{\textbf{$VSM_{Google}$}} & \multicolumn{2}{c}{\textbf{$VSM_{Tencent}$}} \\
\cline{2-11}          & \textbf{AP} & \textbf{MAP} & \textbf{AP} & \textbf{MAP} & \textbf{AP} & \textbf{MAP} & \textbf{AP} & \textbf{MAP} & \textbf{AP} & \textbf{MAP} \\
    \hline
    \textbf{EN-CH Avg. (14)} & 49.04 & 74.16 & 50.18 & 73.74 & 50.56 & 74.13 & 50.81 & 74.24 & \textbf{50.92} & \textbf{74.33} \\
    \textbf{EN-KO Avg. (1)} & 62.14 & 71.31 & 67.99 & 71.33 & 67.95 & \textbf{71.62} & \textbf{68.16} & 70.22 & 65.80  & 70.22 \\
    \textbf{EN-JA Avg. (1)} & \textbf{59.34} & \textbf{78.25} & 49.00    & 72.41 & 49.04 & 72.59 & 51.84 & 73.31 & 52.09 & 72.51 \\
    \textbf{EN-DE Avg. (1)} & 70.83 & 86.20  & 71.38 & 85.94 & 71.41 & 85.95 & \textbf{71.89} & \textbf{88.94} & 71.05 & 86.25 \\
    \textbf{Total Avg. (17)} & 51.70  & \textbf{74.94} & 52.40  & 74.24 & 52.72 & 74.59 & \textbf{53.13} & 74.81 & 53.05 & 74.68 \\
    \hline
    \end{tabular}
  \label{tab:basic_VSM_performance}%
\end{table*}%

\begin{table*}[t]
  \footnotesize
  \centering
  \caption{The score of AP, MAP. We compare $VSM_{Tencent}$+AVIATE with $VSM_{Tencent}$ and $VSM_{Tencent}$+TAROT for RQ1, with $VSM_{Tencent}$+AVIATE+OptAbbr for Additional Observation. ``\textuparrow'' highlights the enhancement brought by AVIATE to $VSM_{Tencent}$.}
    \begin{tabular}{c|
    >{\centering\arraybackslash}p{1cm}>{\centering\arraybackslash}p{1cm}|
    >{\centering\arraybackslash}p{1cm}>{\centering\arraybackslash}p{1cm}|
    cc|
    >{\centering\arraybackslash}p{1cm}>{\centering\arraybackslash}p{1cm}}
    \hline
          & \multicolumn{2}{c|}{\multirow{2}{2cm}{\centering\textbf{$VSM_{Tencent}$}}} 
          & \multicolumn{2}{c|}{\parbox[t]{2cm}{\centering $VSM_{Tencent}$  \\ \textbf{+ TAROT} \vspace{2pt}}} 
          & \multicolumn{2}{c|}{\parbox[t]{2cm}{\centering $VSM_{Tencent}$  \\ \textbf{+ AVIATE} \vspace{2pt}}} 
          & \multicolumn{2}{c}{\parbox[t]{2.4cm}{\centering $VSM_{Tencent}$  \\ \textbf{+ AVIATE + OptAbbr}}} \\
\cline{2-9}          & \textbf{AP} & \textbf{MAP} & \textbf{AP} & \textbf{MAP} & \textbf{AP (\textuparrow)} & \textbf{MAP (\textuparrow)} & \textbf{AP} & \textbf{MAP} \\
    \hline
    Arthas  & 42.67  & 74.16  & 53.96  & 77.21  & 61.65 (+18.98)  & 83.78 (+9.62)  & 61.77  & 83.58  \\
    bk-cmdb & 49.99  & 70.43  & 45.62  & 76.00  & 59.16 (+9.17)  & 83.28 (+12.85)  & 59.09  & 83.37  \\
    Canal & 30.01  & 62.45  & 47.84  & 70.65  & 60.37 (+30.36)  & 74.17 (+11.72)  & 60.19  & 74.06  \\
    Druid & 32.58  & 63.28  & 40.02  & 72.63  & 58.92 (+26.34)  & 76.34 (+13.06)  & 58.88  & 75.96  \\
    Emmagee  & 48.82  & 67.56  & 51.45  & 69.25  & 59.24 (+10.42)  & 77.39 (+9.83)  & 59.50  & 77.69  \\
    Nacos & 36.10  & 63.40  & 34.42  & 62.67  & 47.31 (+11.21)  & 69.26 (+5.86)  & 47.03  & 69.53  \\
    NCNN  & 46.73  & 75.66  & 60.22  & 78.97  & 63.75 (+17.02)  & 83.11 (+7.45)  & 72.61  & 83.08  \\
    Pegasus  & 68.08  & 83.57  & 85.65  & 94.36  & 94.69 (+26.61)  & 96.15 (+12.58)  & 94.64  & 96.05  \\
    QMUI\_Android  & 44.92  & 65.57  & 46.90  & 68.14  & 48.17 (+3.25)  & 64.95 (-0.62)  & 49.41  & 66.75  \\
    QMUI\_IOS & 75.64  & 96.88  & 79.52  & 98.44  & 83.85 (+8.21)  & 98.44 (+1.56)  & 87.91  & 96.88  \\
    Rax   & 53.56  & 73.00  & 69.82  & 85.82  & 86.50 (+32.94)  & 92.45 (+19.45)  & 86.46  & 92.47  \\
    San   & 30.70  & 62.76  & 37.45  & 66.72  & 42.19 (+11.49)  & 67.22 (+4.46)  & 42.29  & 67.10  \\
    Weui  & 76.22  & 89.68  & 76.00  & 90.40  & 82.59 (+6.37)  & 93.27 (+3.59)  & 82.49  & 93.23  \\
    xLua  & 76.92  & 92.15  & 87.78  & 93.91  & 95.04 (+18.12)  & 97.76 (+5.61)  & 95.70  & 97.50  \\
    \textbf{Above 14 Avg. (EN-CH)} & \textbf{50.92} & \textbf{74.33} & \textbf{58.33} & \textbf{78.94} & \textbf{67.39 (+16.46)} & \textbf{82.68 (+8.36)} & \textbf{68.43} & \textbf{82.66} \\
    \hline
    Konlpy (EN-KO) & 65.80  & 70.22  & 73.13  & 76.09  & 76.37 (+10.57)  & 76.93 (+6.71)  & 77.77  & 78.42  \\
    Cica (EN-JA) & 52.09  & 72.51  & 64.46  & 82.76  & 70.23 (+18.14)  & 82.50 (+9.99)  & 70.42  & 82.50  \\
    Awesome-berlin  (EN-DE) & 71.05  & 86.25  & 87.16  & 94.08  & 95.27 (+24.22)  & 94.98 (+8.73)  & 95.45  & 96.33  \\
    \hline
    \textbf{Total 17 Avg.} & \textbf{53.05} & \textbf{74.68} & \textbf{61.26} & \textbf{79.89} & \textbf{69.72 (+16.67)} & \textbf{83.06 (+8.38)} & \textbf{70.68} & \textbf{83.21} \\
    \hline
    \end{tabular}%
  \label{tab:AVIATE_VSM_performance}%
\end{table*}%

For \textbf{RQ1}, we first conducted experiments to identify the optimal IR-based approach with the single translation result.
In alignment with our approach, we employed VSM as the foundational IR model.
We introduced the basic VSM and the VSM based on the single translation result for comparison.
For the former, we directly applied VSM on the artifacts without any translating process, and label this version as $VSM$.
For the latter, we employed four translators (NLLB-1.3B model, M2M-100-12B model, Google Translate, and Tencent Translate, referred to Section~\ref{chap:approach_step2}) to translate the artifacts before applying VSM, and label these versions as $VSM_{NLLB}$, $VSM_{M2M-100}$, $VSM_{Google}$, and $VSM_{Tencent}$.
By comparing them, we can identify an optimal single-translation-based IR approach.
Then, we utilized AVIATE to enhance this best translation results with the consensual biterms extracted from multiple translation variants, and label this version as $VSM_{translator}$+AVIATE, where $translator$ depends on the previous comparison.
By evaluating the performance before and after enhancement, we can assess the improvement brought by AVIATE.
To further validate the advantage of utilizing multiple translation variants in AVIATE, we introduced TAROT~\cite{DBLP:conf/kbse/GaoKSMEMRSZ22}, an existing enhancement strategy for IR-based approaches, which can be utilized to enhance traceability on the translated artifacts. 
We label the optimal single-translation-based approach enhanced by it as $VSM_{translator}$+TAROT.
TAROT also utilizes the consensual biterms to enhance the associations between the artifacts.
However, since TAROT was not designed to address the multilingual issue, it does not consider multiple translation variants during the extraction process.
Besides extracting consensual biterms to enrich the text, TAROT further adjusts the IR value (i.e. the text similarity between artifacts) based on the quality of consensual biterms between the pairs of artifacts (a type of post-processing).
While AVIATE adjusts the weighting factors by modifying the repetition count of consensual biterms based on their distinctiveness (a type of pre-processing).
The advantages of IR-based approaches with machine translation in addressing multilingual artifact traceability has been demonstrated in Lin et al.'s research~\cite{DBLP:journals/ese/LinLC22}.
Gao et al.'s work also showcases that TAROT can effectively enhance traceability between artifacts.
Therefore, the AVIATE's improvement compared to them represents a state-of-the-art advancement for this task.

For \textbf{RQ2}, to evaluate the contributions of AVIATE's different stages (including the use of consensual biterms from multiple translations and the adjustment of weighting factors based on distinctiveness), we introduced a version labeled as $VSM_{translator}$+MultiTV CoB. In this version, we enriched the artifact texts using the consensual biterms extracted from multiple translation variants as described in Section~\ref{chap:approach_step4}. However, each consensual biterm was appended into the artifact only once, without the weighting factor adjustment based on the distinctiveness as discussed in Section~\ref{chap:approach_step5}.
By comparing $VSM_{translator}$ with $VSM_{translator}$+MultiTV CoB, we can observe the contributions of the multiple-translation-augmented consensual biterms. By comparing $VSM_{translator}$+MultiTV CoB with the complete $VSM_{translator}$+AVIATE, we can further observe the contribution of adjusting their weighting factors.


\section{Result And Discussion}

\begin{itemize}[label=$\bullet$, left=0cm, topsep=0pt]
    \item \textbf{RQ1: How much improvement can AVIATE bring to the single-translation-based IR approach?}
\end{itemize}
Table~\ref{tab:basic_VSM_performance} shows the performance of the basic VSM and four machine translation-based VSM.
We observe that the translation-based VSM consistently outperforms the basic VSM in terms of average AP performance: $VSM_{NLLB}$ improves by 0.70 points, $VSM_{M2M-100}$ by 1.02, $VSM_{Google}$ by 1.43, and $VSM_{Tencent}$ by 1.35 across the 17 projects.
This reconfirms the effectiveness of machine translation to improve the IR-based traceability on the multilingual artifacts.
However, the highest average MAP performance is achieved by the basic VSM, without any translation.
This is due to the poor performance of translators in English-Japanese translation, which significantly lowers the overall performance of the translation-based VSMs, particularly for the NLLB-1.3B and M2M-100-12B models. 
The partial translation failures, where Japanese content remains untranslated within the translation results, introduce substantial noise into the IR calculations, thereby reducing the VSM's effectiveness. 
This highlights the crucial role of the translation quality in translation-based IR approaches.
In summary, considering that ``English-Chinese'' are the main cases of the multilingual artifacts, we select $VSM_{Tencent}$, which shows the best AP and MAP performance on 14 English-Chinese projects, as the optimal single-translation-based approach.
Its overall performance across the 17 projects is only slightly below that of the best-performing $VSM_{Google}$ by approximately 0.1.
Next, we will evaluate the improvement of AVIATE based on it.

Table~\ref{tab:AVIATE_VSM_performance} shows the performance of $VSM_{Tencent}$, $VSM_{Tencent}$+ TAROT, and $VSM_{Tencent}$+AVIATE.
Comparing $VSM_{Tencent}$ with $VSM_{Tencent}$+TAROT, we observe that TAROT can effectively enhance the performance of $VSM_{Tencent}$.
It brings an average increase of 8.21 points (a 15.47\% increase) in AP and 5.21 (6.98\%) in MAP across 17 projects.
However, AVIATE can offer an even greater improvement compared to it.
In terms of AP, AVIATE provides an average improvement of 16.67 (31.43\%) compared to $VSM_{Tencent}$, with a maximum increase of 30.36 (101.17\%) observed in the Canal project.
Regarding MAP, AVIATE results in an average improvement of 8.38 (11.22\%), with a maximum increase of 19.45 (26.64\%) observed in the Rax project.
Therefore, we argue that compared to the conventional enhancement strategy TAROT, AVIATE's strategy of extracting consensual biterms from multiple translation variants and emphasizing the critical sections by adjusting weighting factors based on the distinctiveness can more effectively enhance the traceability performance of the single-translation-based IR approach on multilingual artifacts.
Moreover, AVIATE demonstrates consistent performance improvements across various languages, including English-Chinese, English-Korean, English-Japanese,  and English-German.
This consistent performance suggests a promising prospect for the broader application of AVIATE.

To better understand the advantages of AVIATE, we conducted several case studies on it.
First, we argue that one of the advantages of AVIATE is its ability to extract consensual biterms from multiple translation variants, which increases the likelihood of identifying potential links between artifacts. 
This has been showcased in the motivation example in Section~\ref{chap:motivating_example}. 
In this example, even using the optimal Tencent translator, we could not extract the critical consensual biterm ``attributedisabled'' solely from its translation results. 
Because the Tencent translated the commit as ``... attributes such as disabled...'', where the relationship between two critical terms ``attribute'' and ``disabled'' is the exemplification, which does not align with the modifying relationship present in the issue.
However, by incorporating the multiple translations, AVIATE could successfully recall this consensual biterm through the Tencent translation of the issue and the NLLB translation of the commit, thereby confirming the traceability link between the artifacts.
Further statistical analysis also supports this advantage.
We found that the single-translation-based TAROT could extract an average of 4773 consensual biterms across the 17 projects, while the multi-translation-based AVIATE could extract an average of 5466 consensual biterms.
This increase in the number of extracted consensual biterms underscores AVIATE's enhanced capability in uncovering the potential traceability associations between artifacts.
Another advantage of AVIATE is its adeptness in effectively highlighting important information.
Taking a pair of linked artifacts (Issue 234, Commit e615c06d...) in the San project as an example, TAROT and AVIATE almost select the same consensual biterms for them. 
TAROT enhances their association by applying a coefficient of 2.507 to the original IR value, which is determined based on the consensual biterms between them.
Under this enhancing method, this link ranks 837th in the overall ranking.
In contrast, AVIATE elevates this link to the 38th by emphasizing their shared consensual biterms ``fixmethod'' and ``exportmethod'' within the commit message, appending them 59 times to the commit text based on the average distinctiveness of the commit message section.
This notable rank improvement underscores the effectiveness of the distinctiveness-based pre-processing emphasis strategy within AVIATE.
In the next RQ2, we will analyze the specific contributions of the above two advantageous strategies.

\begin{table}[t]
  \footnotesize
  \centering
  \caption{The score of computed AP, MAP evaluating the approaches ($VSM_{Tencent}$, $VSM_{Tencent}$+MultiTV CoB, and $VSM_{Tencent}$+AVIATE) (RQ2).}
    \resizebox{0.47\textwidth}{!}{\begin{tabular}{c|cc|cc|cc}
    \hline
    & \multicolumn{2}{c|}{\multirow{2}{1.8cm}{\centering\textbf{$VSM_{Tencent}$}}}
    & \multicolumn{2}{c|}{\parbox[t]{1.8cm}{\centering $VSM_{Tencent}$  \\ \textbf{+ MultiTV CoB} \vspace{2pt}}}
    & \multicolumn{2}{c}{\parbox[t]{1.8cm}{\centering $VSM_{Tencent}$  \\ \textbf{+ AVIATE} \vspace{2pt}}} \\
\cline{2-7}          & \textbf{AP} & \textbf{MAP} & \textbf{AP} & \textbf{MAP} & \textbf{AP} & \textbf{MAP}  \\
    \hline
    Arthas  & 42.67  & 74.16  & 46.98  & 77.59  & 61.65  & 83.78    \\
    bk-cmdb & 49.99  & 70.43  & 52.77  & 75.84  & 59.16  & 83.28    \\
    Canal & 30.01  & 62.45  & 37.38  & 69.25  & 60.37  & 74.17    \\
    Druid & 32.58  & 63.28  & 35.40  & 67.54  & 58.92  & 76.34    \\
    Emmagee  & 48.82  & 67.56  & 51.21  & 68.13  & 59.24  & 77.39    \\
    Nacos & 36.10  & 63.40  & 36.01  & 63.39  & 47.31  & 69.26    \\
    NCNN  & 46.73  & 75.66  & 48.51  & 77.71  & 63.75  & 83.11    \\
    Pegasus  & 68.08  & 83.57  & 74.44  & 89.16  & 94.69  & 96.15    \\
    QMUI\_Android  & 44.92  & 65.57  & 47.22  & 66.75  & 48.17  & 64.95    \\
    QMUI\_IOS & 75.64  & 96.88  & 77.59  & 96.88  & 83.85  & 98.44    \\
    Rax   & 53.56  & 73.00  & 63.67  & 78.64  & 86.50  & 92.45    \\
    San   & 30.70  & 62.76  & 34.44  & 65.52  & 42.19  & 67.22    \\
    Weui  & 76.22  & 89.68  & 77.88  & 90.69  & 82.59  & 93.27    \\
    xLua  & 76.92  & 92.15  & 87.32  & 95.83  & 95.04  & 97.76    \\
    \textbf{Above 14 Avg. (EN-CH)} & \textbf{50.92} & \textbf{74.33} & \textbf{55.06} & \textbf{77.35} & \textbf{67.39} & \textbf{82.68}  \\
    \hline
    Konlpy (EN-KO) & 65.80  & 70.22  & 70.38  & 71.79  & 76.37  & 76.93    \\
    Cica (EN-JA) & 52.09  & 72.51  & 59.52  & 77.84  & 70.23  & 82.50    \\
    Awesome-berlin  (EN-DE) & 71.05  & 86.25  & 79.28  & 92.05  & 95.27  & 94.98   \\
    \hline
    \textbf{Total 17 Avg.} & \textbf{53.05} & \textbf{74.68} & \textbf{57.65} & \textbf{77.92} & \textbf{69.72} & \textbf{83.06}  \\
    \hline
    \end{tabular}}%
  \label{tab:exp_for_RQ2&3}%
\end{table}%

Additionally, we present an error case analysis to illustrate the potential for further improving AVIATE in future work.
We identify a case in San project where Issue 364 has a true trace link with Commit 39fdbda3..., but our approach fails to recover the link between them.
This issue briefly describes the bug as ``\begin{CJK*}{UTF8}{gbsn}san.parseTemplate解析问题\end{CJK*}'' (A parsing bug from san.parseTemplate), while this commit also just provides a concise description about the cause of the issue, stating ``\begin{CJK*}{UTF8}{gbsn}闭合标签有属性。close tag has attributes\end{CJK*}'' (same meaning for the two clauses in different languages).
Although our approach is able to extract ``closetag'' as a distinctive consensual biterm representing the related system functionalities across the whole project, the issue and the commit still have few overlapped terms.
This lack of shared terms results in a low textual similarity between the artifacts, ultimately leading to the failure in detecting their traceability relationships.
This case represents a common challenge that may arise in practical traceability recovery: issues and commits are likely to be too briefly written during the intense development. 
To address this challenge, we propose to explore additional information for reference, such as version history or reporter information (e.g., components the issue reporter frequently focuses on), as suggested by Wang et al.\cite{DBLP:journals/smr/WangL16}.
Incorporating these additional references may enhance AVIATE's ability to recover trace links when the textual information of the artifacts alone is not sufficient.

\vspace{1em}
\begin{itemize}[label=$\bullet$, left=0cm, topsep=0pt]
    \item \textbf{RQ2: How much contribution do the different strategy settings within AVIATE make?}
\end{itemize}
Table~\ref{tab:exp_for_RQ2&3} presents the experiments for RQ2.
Comparing $VSM_{Tencent}$ and $VSM_{Tencent}$+MultiTV CoB, we observe that incorporating consensual biterms from multiple translation variants to enrich artifact texts has shown a promising improvement for the single-translation-based IR approaches.
$VSM_{Tencent}$+MultiTV CoB exhibited an average AP increase of 4.60 (8.66\%) and an average MAP increase of 3.24 (4.34\%) over $VSM_{Tencent}$ across 17 projects.
Furthermore, emphasizing the consensual biterms from summary information (i.e., the issue summary and the commit message) based on their distinctiveness significantly amplifies this improvement, which lead to an additional average AP increase of 12.08 (20.95\%) and 5.14 (6.60\%) for MAP (comparing $VSM_{Tencent}$+AVIATE to $VSM_{Tencent}$+MultiTV CoB).
Therefore, we consider that each strategy within AVIATE contributes to better handling multilingual artifact traceability.

\begin{table}[t]
  \footnotesize
  \centering
  \caption{The foreign term percentage (reported by Lin et al.~\cite{DBLP:journals/ese/LinLC22}), $SummaryDetailRatio$ (calculated as the total word number ratio of issue summaries and commit messages to issue descriptions and commit diffs), and the percentage of abbreviations for bilingual projects.}
    \resizebox{0.47\textwidth}{!}{\begin{tabular}{c|c|c|c}
    \hline
     & \textbf{ForeignTerms} & \textbf{SummaryDetailRatio} & \textbf{Total Abbr.} \\
    \hline
    Arthas & 14.60\% & 3.58\% & 1.10\% \\
    bk-cmdb & 8.30\% & 1.55\% & 2.27\% \\
    Canal & 5.40\% & 1.32\% & 2.17\% \\
    Druid & 7.30\% & 1.13\% & 2.73\% \\
    Emmagee  & 21.90\% & 8.26\% & 1.16\% \\
    Nacos & 1.00\% & 0.27\% & 1.89\% \\
    NCNN & 28.00\% & 3.11\% & 0.78\% \\
    Pegasus & 35.30\% & 1.98\% & 1.57\% \\
    QMUI\_Android & 16.80\% & 4.52\% & 6.71\% \\
    QMUI\_IOS & 20.80\% & 9.45\% & 9.32\% \\
    Rax & 9.00\% & 1.83\% & 2.14\% \\
    San & 8.20\% & 2.05\% & 2.48\% \\
    Weui & 7.10\% & 2.82\% & 1.04\% \\
    xLua & 29.50\% & 1.13\% & 1.83\% \\
    Konlpy & 7.00\% & 4.13\% & 1.31\% \\
    Cica & 11.70\% & 2.49\% & 0.75\% \\
    Awesome-berlin & 31.00\% & 3.39\% & 0.92\% \\
    \hline
    \end{tabular}}%
  \label{tab:statistics}%
\end{table}%

Specifically, we observed that the improvement achieved by simply introducing consensual biterms from multiple translations (without emphasizing them based on their distinctiveness) appears to be limited, with average increase of 4.60 and 3.24 in AP and MAP, compared to the increase of 16.67 and 8.38 for the complete AVIATE.
We attribute this limitation to the relatively low proportion of foreign terms in the current dataset, which restricts the potential of AVIATE, a strategy designed to address multilingual issues in artifact traceability.
Table~\ref{tab:statistics} shows the percentage of foreign (non-English) terms in various projects within the dataset, ranging from 1\% to 35.3\%.
We conducted a Pearson correlation analysis~\cite{cohen2009pearson} between the improvement of AVIATE (highlighted by ``\textuparrow'' in Table~\ref{tab:AVIATE_VSM_performance}) and the foreign term percentage of projects.
The analysis revealed a weak positive correlation (+0.15) between the improvement in AP and the foreign term percentage, which further supports our hypothesis.
We speculate that in open-source communities, users are more inclined to communicate in English, resulting in relatively fewer instances of mixed languages in this GitHub-obtained dataset.
Conversely, in real-world closed industrial development environments, developers may more frequently use their native language (non-English) for communication. 
Therefore, in future studies, we aim to evaluate AVIATE on industrial datasets to validate its practical effectiveness in environments with higher proportions of foreign terms.

To amplify the effect of consensual biterms from multiple translations and emphasize the concise yet important summary information, we adjust the weighting factor of the consensual biterms in issue summaries and commit messages based on their distinctiveness. 
This strategy has shown a significant enhancement in the performance of AVIATE in previous analyses.
We analyzed the improvement brought by AVIATE in relation to the ratio of summary information to detailed descriptions (i.e. $SummaryDetailRatio$ in Table~\ref{tab:statistics}).
We found that the AVIATE’s improvements in AP and MAP show a moderate negative correlation with $SummaryDetailRatio$ (-0.44 and -0.38, respectively).
This indicates that AVIATE can better handle the artifacts that have an uneven distribution of information, particularly those where the summary information is relatively scarce compared to detailed descriptions.

\vspace{1em}
\begin{itemize}[label=$\bullet$, left=0cm, topsep=0pt]
    \item \textbf{Additional Observation: Expanding abbreviations}
\end{itemize}
The data analysis in RQ2 indicates that the characteristics of artifacts can influence the effectiveness of AVIATE.
To further enhance the quality of artifact texts for better traceability, we expanded abbreviations within the artifacts, referring to the method proposed by Jiang et al.~\cite{DBLP:journals/tse/Jiang0Z020}. 
We use the \texttt{corpus.words} package in NLTK~\cite{NLTK} to verify whether a term is a complete English word or an abbreviation. 
For the abbreviations, we select appropriate full forms for them based on their context within the artifact.

Table~\ref{tab:AVIATE_VSM_performance} reports the results of $VSM_{Tencent}$+AVIATE enhanced with abbreviation expansion (denoted by ``+OptAbbr’’). We found that although the improvement is not significant, abbreviation expansion can enhance traceability performance to some extent, with an average increase of 0.96 (1.38\%) in AP and 0.15 (0.18\%) in MAP. The limited impact of abbreviation expansion is likely due to the low frequency of abbreviations within the projects, ranging from 0.78\% to 9.32\% (average 2.36\%). This low proportion constrains the potential for improvement.
However, we observed that despite the NCNN project having only 0.78\% abbreviations, the abbreviation expansion significantly improved its AP by 8.86. Upon further analysis of this project, we found that the abbreviation expansion effectively eliminated the redundant consensual biterms. Specifically, before applying abbreviation expansion, we extracted a total of 4099 consensual biterms from the NCNN’s issues and commits. After applying abbreviation expansion, this number decreased to 3607, indicating that 492 redundant consensual biterms were removed (e.g. ``lens'', ``lenstr'' -> ``lenstring''). This finding suggests that abbreviation expansion not only improves the likelihood of identifying consensual biterms by completing abbreviations, but also enhances their overall quality by eliminating the scrappy one.
Therefore, we consider abbreviation expansion as an effective strategy as well.

\section{Related Work}
In this section, we introduces two main technologies for the traceability recovery, i.e. information retrieval (IR) and machine learning (ML),
and reviews the existing explorations towards the multilingual issue in the software development.

\textbf{IR-based traceability recovery and enhancements:}
The IR techniques have been widely applied in the automated traceability recovery~\cite{DBLP:conf/icse/Cleland-HuangGHMZ14}. 
They quantify and rank the possible traceability relationships between the pairs of artifacts (e.g., issues and commits) by calculating the textual similarity between them.
The Vector Space Model (VSM)~\cite{DBLP:journals/tse/AntoniolCCLM02}, Latent Semantic Indexing (LSI)~\cite{DBLP:conf/icse/MarcusM03}, and Jenson-Shannon model (JS)~\cite{DBLP:conf/iwpc/AbadiNS08} are all commonly used IR models in this field.
Among the discussed IR models, VSM, despite its simplicity, often exhibits advantages in various tracing scenarios~\cite{DBLP:conf/sigsoft/LoharAZC13}, including in Lin et al.'s exploration~\cite{DBLP:journals/ese/LinLC22} on traceability recovery on bilingual projects.
Therefore, we choose it as the foundational IR model in our work.
Two main kinds of enhancing strategies have also been proposed to improve IR-based approaches.
One focuses on emphasizing the relatively important information by reducing the ``noisy contents’’ in artifacts, such as removing the semantically duplicate information~\cite{DBLP:conf/iwpc/LuciaPOPP11} or filtering insignificant terms based on their POS tags~\cite{DBLP:journals/smr/CapobiancoLOPP13,DBLP:journals/infsof/AliCHH19}.
The other chooses to enrich the artifacts by supplementing additional information, such as expanding abbreviations into complete terms~\cite{DBLP:journals/tse/Jiang0Z020}, or using the consensual biterms to enhance the semantic connections between artifacts~\cite{DBLP:conf/kbse/GaoKSMEMRSZ22,gao2024triad}.
However, most of these strategies do not consider the semantic gaps introduced by multilingualism.
Thus, we use consensual biterms across translation variants to enrich the artifacts and thus improve IR-based traceability recovery on bilingual software projects.

\textbf{ML-based traceability recovery:}
Recently, a growing body of studies have proposed using machine learning techniques for the traceability recovery.
Mills et al.~\cite{DBLP:conf/icsm/MillsEH18} transformed the tracking relationship into a binary classification issue, i.e. whether the link between pairs of artifacts is valid or not. They used machine learning algorithms to extract features from existing trace links and infer the validity of artifact links.
Guo et al.~\cite{DBLP:conf/icse/0004CC17} employed the RNN units to map the artifacts into semantic vectors, and used semantic relation evaluation layers to predict the probability of links between them.
Lin et al.~\cite{DBLP:conf/icse/LinLZ0C21} evaluated three different BERT-based frameworks to predict whether artifacts are related.
The use of learning-based approaches relies on the accumulation of pre-existing trace links.
Although some efforts have been made to reduce the requirement for training data, such as active learning~\cite{DBLP:journals/jzusc/DuSHYW20} and transfer learning~\cite{DBLP:journals/corr/abs-2207-01084}, the IR-based approaches are still more practical compared to them, because the IR-based approaches can function effectively without any pre-existing trace links, which makes them more suitable for the real-world development.

\textbf{Multilingualism in the software development and traceability:}
In today’s global software development, machine translation has been widely employed as a solution for the multilingual issue.
For example, Calefato et al.~\cite{DBLP:conf/icgse/CalefatoLM10} applied machine translation in synchronous, text-based chats to overcome language barriers during distributed requirements engineering workshops, while Wouters et al.~\cite{DBLP:conf/cscwd/WoutersKS13} maintained a domain ontology through the machine translation to facilitate collaborative design activities among experts from diverse languages and domains.
These works all focus on bridging communication gaps by converting between different languages, rather than the challenges posed by multilingualism inside different kinds of artifacts (our main focus of this paper).
In particular, the work by Xu et al.~\cite{DBLP:journals/ese/XuXXLL18} is more closely related to ours.
They proposed a cross-language search approach to assist Chinese developers in finding solutions on Stack Overflow.
Their approach extracts both Chinese and English keywords from Chinese Java questions. For Chinese keywords, they employ machine translation and domain-specific vocabularies to translate them into appropriate English terms, and ultimately select the most relevant keywords for searching solutions on Stack Overflow.
Both of our approaches leverage machine translation to help IR systems better identify linked content.
The domain-specific translation used in Xu et al.’s work may also help us improve AVIATE.
Meanwhile, to address multilingualism in traceability recovery, Lin et al.~\cite{DBLP:journals/ese/LinLC22} compared the effectiveness of IR models on machine-translated artifacts with a deep-learning approach based on BERT. Their experiments on 17 bilingual projects showed that VSM-based models outperformed both other IR models and the BERT-based approach.
Inspired by their work, we conduct our research to further improve traceability recovery on bilingual projects.
The key advantages of our approach are: 1) we introduce multiple translators to minimize the potential vocabulary mismatch problems caused by multilingualism; 2)  instead of directly using the translation variants (e.g., concatenating them with the original texts), we enrich the artifacts with the consensual biterms across translation variants, which helps reduce redundant information and better emphasize the connections between artifacts; 3) unlike Lin et al. treating all sections of the artifact equally, we adjust the weighting factors to ensure that important information within critical sections receives adequate attention in the IR model. The evaluation results based on our experiments demonstrate the effectiveness of our proposed approach.

\section{Threats To Validity}
\textbf{Internal Threats.}
A potential internal threat to our approach is the quality of the translators used. Since AVIATE extracts consensual biterms from the translation results of artifacts, the quality of translations may affect the quality of the extracted biterms. However, the four translators we used have been applied in various researches~\cite{DBLP:journals/corr/abs-2304-04675, DBLP:journals/corr/WuSCLNMKCGMKSJL16}, demonstrating their reliability. Additionally, by focusing on biterms that have achieved consensus across multiple translation variants, we minimize the impact of any single translator's quality, ensuring the stability and reliability of the extraction results.
Another potential internal threat is the quality of natural language analysis provided by CoreNLP. We cannot guarantee the complete accuracy of CoreNLP's POS tagging, and dependency parsing. Nonetheless, research~\cite{DBLP:journals/infsof/AliCHH19} has shown that the existing NLP tools can perform reliably on well-structured texts, and we have not observed significant errors in our analyses. Therefore, we believe this potential internal threat is acceptable.

\textbf{External Threats.}
There are several external threats to consider in our study. Firstly, while attempting to replicate the translation-based VSM as described by Lin et al.~\cite{DBLP:journals/ese/LinLC22}, we manually reproduced the model and reorganized the dataset for better application. However, the lack of detailed implementation code and specific project performance (only providing the average data) for the VSM model in their paper might lead to slight discrepancies between our results and theirs. To address this concern, we have made our code and reorganized dataset publicly available to facilitate the validation of our research outcomes. Nevertheless, these differences do not undermine our observations regarding the enhancement brought by AVIATE.
Secondly, our focus solely on VSM, without exploring other IR models, might be perceived as a limitation. However, since both VSM and other IR models operate on similar principles, i.e., ranking documents based on similarity between queries and documents, our observations regarding AVIATE's effectiveness in VSM are likely applicable to other IR models as well.
Thirdly, the construction process of the dataset we referenced indicates a higher trace link density compared to real-world project data.
This might slightly inflate the performance metrics reported in this study.
To address this concern, we plan to evaluate the approach performance on more realistic multilingual datasets in the future.
Lastly, we did not incorporate large language models (LLMs) in this study because: (1) Zhu et al.~\cite{DBLP:journals/corr/abs-2304-04675} reported that LLMs did not achieve the best performance in machine translation, especially for multilingual translation tasks; (2) Alberto et al.~\cite{DBLP:conf/re/RodriguezDC23} reported that prompt learning via LLMs (without pre-defined traces) are so-far not stable and effective enough to directly perform traceabibility recovery.
However, we still expect to use the LLMs' potentials to enrich contexts for mulilingual traceability recovery.

\section{Conclusion And Future Works}

To address the challenges posed by multilingualism in IR-based traceability recovery, we propose an enhancement strategy, named AVIATE.
It first extracts consensual biterms from the multiple translation variants of the artifacts, then employs a TF-IDF-like metric to measure and select distinctive consensual biterms to enrich the artifact text.
Experiments on 17 bilingual projects has demonstrated its effectiveness.
In future work, we plan to use domain-specific translations to improve AVIATE's performance on specific projects.
We also plan to bring in more information such as history versions to issues and commit messages before we apply AVIATE to trace them (e.g., consulting the work of Wang et al.\cite{DBLP:journals/smr/WangL16}).
Finally, we expect to contact the companies involved in the bilingual dataset to explore how the performance of automated traceability approaches is acceptable for bilingual projects in practice.


\begin{acks}

This work is supported by National Natural Science Foundation of China (No.62025202, No.62072227, No.62202219, No.62302210, No.72371125), Jiangsu Provincial Key Research and Development Program (No.BE20210022), Natural Science Foundation of Jiangsu Province (No.BK20241195), and Innovation Project and Overseas Open Project of State Key Laboratory for Novel Software Technology (Nanjing University) (ZZKT2024A18, ZZKT2024B07, KFKT2023-A09, KFKT2023A10, KFKT2024A02, KFKT2024A13, KFKT2024A14).
\end{acks}

\bibliographystyle{ACM-Reference-Format}
\bibliography{sample-base}










\end{document}